\documentclass[final,3p,times]{elsarticle}
\usepackage{graphicx}
\usepackage{amssymb}
\usepackage{multirow}

\journal{Elsevier}

\begin{document}

\begin{frontmatter}



\title{Individual popularity and activity in online social systems}
\author[rvt]{Haibo Hu}
\ead{hbh@sjtu.edu.cn}
\author[focal]{Dingyi Han}
\author[rvt]{Xiaofan Wang}

%\cortext[cor1]{Corresponding author. Tel.: +86 21 34203083}

\address[rvt]{Complex Networks and Control Lab, Shanghai Jiao Tong University, Shanghai 200240,
China}
\address[focal]{Apex Data and Knowledge Management Lab, Shanghai Jiao Tong University, Shanghai
200240, China}

\begin{abstract}
We propose a stochastic model of web user behaviors in online social
systems, and study the influence of attraction kernel on statistical
property of user or item occurrence. Combining the different growth
patterns of new entities and attraction patterns of old ones,
different heavy-tailed distributions for popularity and activity
which have been observed in real life, can be obtained. From a
broader perspective, we explore the underlying principle governing
the statistical feature of individual popularity and activity in
online social systems and point out the potential simple mechanism
underlying the complex dynamics of the systems.
\end{abstract}


\end{frontmatter}


%%\newpage
\section{Introduction}
Currently the WWW is undergoing a landmark revolution from the
traditional Web 1.0 to Web 2.0 characterized by social collaborative
technologies, such as social networking site, blog, Wiki and
folksonomy. The social Web (or more specifically, online social
systems), a label that includes both social networking sites (such
as $MySpace$ and $Facebook$) and social media sites (such as $Digg$,
$CiteULike$ and $Flickr$), is changing the way content is created
and distributed. Web-based authoring tools enable users to rapidly
publish content, from stories and opinion pieces on weblogs, to
photographs and videos on $Flickr$ and $YouTube$, to advice on
$Yahoo! Answers$, and to web discoveries on $Del.icio.us$ and
$Furl$. The availability of large-scale electronic databases has
delivered us extraordinary new insights on the human behaviors and
human dynamics on the web. The clear patterns and regularities in
individual distributions in respect of popularity and activity in
some online social systems have been revealed [1-8].

Evidently web users vary widely in their activity levels. Take
$Digg$ as example, some users casually browse the front page, voting
on one or two stories. Others spend hours a day combing the web for
new stories to submit, and voting on stories they found on $Digg$.
Also different items on the web vary widely in their popularity.
Some stories can attract large attention and their influence can
last for a long time while most stories only can attract very little
attention and their impact vanishes rapidly. In social media sites a
$tag$-$cloud$ is usually used to visualize the popularity of items,
or more specifically, tags. Typical tag-clouds have between 30 and
150 tags. The popularity is represented using font sizes, colors or
other visual clues. Fig. 1 shows a tag-cloud with terms related to
Web 2.0.

\begin{figure}
  \centerline{\includegraphics[width=3in]{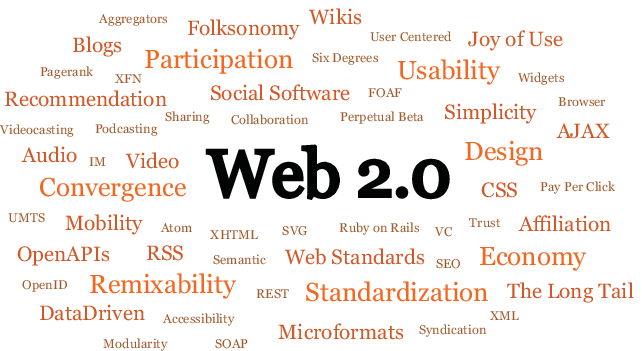}}
  \caption{A tag-cloud with terms related to Web 2.0
(http://en.wikipedia.org/wiki/Tag\_cloud). Generally the font size
of each tag is proportional to the logarithm of its frequency of
appearance within the folksonomy.}
\end{figure}

Recently much attention has been devoted to investigating the
statistical feature of individual popularity and activity in online
social systems. Their distributions show the wide-spread believed
power law or general heavy-tailed ones intermediate between
exponential and power law, such as stretched exponential or
log-normal [2, 5, 9]. Despite the great progress made, little work
is done on the underlying mechanism governing the statistical
feature of  popularity and activity in online social systems, which
will be explored in the work.

We can start our analysis from a time-ordered table of item
assignments. For the system as a whole, we can define an intrinsic
time $T$ as the index of an item assignment into such a table, so
that $T$ runs from 1 to the number of total item assignments. The
temporal process shown in Fig. 2 can be regarded as the process of
appearance of entity $u_i$ or $m_i$. And the frequency of occurrence
for some user/item in the total $T$ events can be defined as its
activity/popularity. Thus activity measures how frequently a user
performs a specific action, such as listening to music, seeing
films, browsing posts and sending friendship invitations to other
users on the web, and popularity measures how frequently an item
(such as music, films, posts and tags) is visited by web users. Note
that for items we can only measure their popularity, while for
users, in some cases, we can measure not only their activity but
also popularity. For instance in online social networks, users can
invite other users to be their friends. Thus we can measure the
activity of users in terms of the number of sent invitations, and
can also measure the popularity of users in terms of the number of
received invitations.

\begin{figure}
  \centerline{\includegraphics[width=3in]{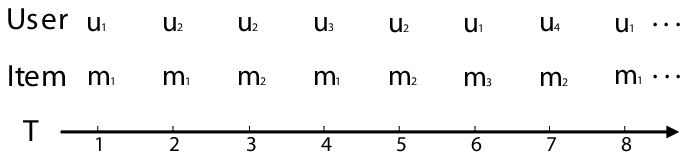}}
  \caption{A schematic illustration of online social systems. It is a
sequence of chronologically ordered users and items.}
\end{figure}

It is natural that the number of distinct items or users $N$
increases with $T$, however different growth patterns can appear.
Generally $N(T) \propto T^\gamma  $, where $\gamma<1$ implies
sub-linear growth while $\gamma=1$ linear growth, i.e. the
generation probability of new individuals is a constant value
(homogeneous Poisson process). Besides the more frequently an
individual appears, the more possibly the individual will appear
once again. Specifically, when an old individual joins the sequence,
the probability that it will be a specific old individual $i$ with
previous frequency of appearance $k_i$ is $\prod {(k_i )}  \propto
k_i^\beta$ ($0 \le \beta  \le 1$). The preference metric $0 \le
\beta  < 1$ implies sub-linear preference while $\beta=1$ linear
preference. The case where $N(T) \propto T$ and $\beta=1$
corresponds to the classic Simon model [10, 11].

When an old individual joins the sequence, the probability that
individual $i$ with frequency $k_i$ is selected can be expressed
as $\prod (k_i ) = k_i^\beta  /\sum\nolimits_j {k_j^\beta  } $.
Thus we can compute the probability $\prod (k)$ that an old
individual of frequency $k$ is chosen, and it is normalized by the
number of individuals of frequency $k$ that exist just before this
step [12, 13]: $\prod {(k) = \sum\nolimits_t {[e_t  = v \wedge k_v
(t - 1) = k]} } /\sum\nolimits_t {|\{u}:k_u (t - 1) = k\} | \sim
k^\beta$, where $e_t  = v \wedge k_v (t - 1) = k$ represents that
at time $t$ the old individual whose frequency is $k$ at time
$t-1$ is chosen. We use $[ \cdot ]$ to denote a predicate (take
value of 1 if expression is true, else 0). Generally $\prod (k)$
has significant fluctuations, particularly for large $k$. To
reduce the noise level, instead of $\prod (k)$ we can study the
cumulative function to obtain the preference metric $\beta$:
$\kappa (k) = \int_0^k {\prod (k)} {\rm{d}}k \sim k^{\beta + 1} $.

\section{Model}
Consider that users are listening to music. At a discrete time step
$T$, a new user may appear with probability $\alpha $, whereas with
probability $1-\alpha $ an existing old user can appear. We can
apply the mean field method to analytically obtain the probability
distribution for individual popularity and activity. When $0 < \beta
< 1$, we have
\begin{equation}
  \frac{{\partial k_i }}{{\partial t}} = (1 - \alpha )\frac{{k_i^\beta  }}{{\sum\nolimits_j {k_j^\beta  }
  }}.
  \label{1}
\end{equation}
According to
\begin{equation}
  \left\{ \begin{array}{l}
 \sum\nolimits_j {k_j^0 }  = t\alpha    \\
 \sum\nolimits_j {k_j^1 }  = t   \\
 \end{array} \right.
 \label{2}
\end{equation}
and
\begin{equation}
  \sum\nolimits_j {k_j^\beta  }  = \mu t\;\;(\alpha  < \mu  < 1),
  \label{3}
\end{equation}
We obtain
\begin{equation}
  \frac{{\partial k_i }}{{\partial t}} = (1 - \alpha )\frac{{k_i^\beta  }}{{\mu
  t}}.
  \label{4}
\end{equation}
The initial condition is $k_i (t_i ) = 1$, where $t_i$ is the time
when the individual $i$ appeared for the first time. Thus
\begin{eqnarray}
 k_i^{1 - \beta } (t) &= & 1 + \ln \left( {\frac{t}{{t_i }}} \right)^{(1 - \beta )\left( {\frac{{1 - \alpha }}{\mu }} \right)} \nonumber \\
  & \sim & (1 - \beta )\left( {\frac{{1 - \alpha }}{\mu }} \right)\ln t
   \label{5}
\end{eqnarray}
and
\begin{equation}
  P(k_i (t) < k) = P\left[ {t_i  > t \cdot \exp \left( { - \frac{{k^{1 - \beta }  - 1}}{{1 - \beta }} \cdot \frac{\mu }{{1 - \alpha }}} \right)}
  \right].
  \label{6}
\end{equation}
The probability density function for $t_i$ is $P_i (t_i ) = \alpha
/(1 + t)$ and thus
\begin{eqnarray}
 P(k_i (t) < k) &=& 1 - P\left[ {t_i  \le t \cdot \exp \left( { - \frac{{k^{1 - \beta }  - 1}}{{1 - \beta }} \cdot \frac{\mu }{{1 - \alpha }}} \right)} \right] \nonumber \\
  &=& 1 - \frac{{t\alpha }}{{t + 1}}\exp \left( { - \frac{{k^{1 - \beta }  - 1}}{{1 - \beta }} \cdot \frac{\mu }{{1 - \alpha }}}
  \right).
  \label{7}
\end{eqnarray}
The probability distribution $P(k)$ for individual popularity and
activity is
\begin{eqnarray}
 P(k)  &=& \frac{{\partial P(k_i (t) < k)}}{{\partial k}} \nonumber \\
  &=& \frac{{t\alpha }}{{1 + t}}\frac{\mu }{{1 - \alpha }} \cdot k^{ - \beta }  \cdot \exp \left( { - \frac{{k^{1 - \beta }  - 1}}{{1 - \beta }} \cdot \frac{\mu }{{1 - \alpha }}}
  \right),
  \label{8}
\end{eqnarray}
which is a stretched exponential distribution. Its complementary
cumulative distribution functions (CCDF) is $P_{\rm{c}} (k) = \exp [
- (k/k_0 )^c ]$, where $c=1-\beta$ and $k_0$ is a constant. When
$\beta \to 1$, $\mathop {\lim }\limits_{\beta \to 1} \frac{{k^{1 -
\beta } - 1}}{{1 - \beta }} = \ln k$, $\mu  = 1$, and
\begin{equation}
  P(k)  = \frac{{t\alpha }}{{(1 + t)(1 - \alpha )}} \cdot k^{ - \left( {1 + \frac{1}{{1 - \alpha }}}
  \right)},
  \label{9}
\end{equation}
which is a power law distribution. Its CCDF is $P_{\rm{c}} (k) \sim
k^{-\nu} $, where $\nu=1/(1-\alpha)$. The special situation of
absent preference $\beta=0$ reduces Eq. (8) to an exponential
distribution. Generally the stretched exponential distribution is
correlative with sub-linear preference while power law distribution
linear preference [14, 15].

\section{Results and discussion}
We ground our empirical analysis on actual log data extracted from
an online media site
$Comic$\footnote{http://comic.sjtu.edu.cn/music.asp} [5] and an
online social network $Wealink$\footnote{http://www.wealink.com}
[6]. Note that our approach of investigation is also applicable to
other online social systems. $Comic$ is located in a large Chinese
university with more than 40,000 undergraduate and graduate
students, and only is accessible to the IP addresses within the
university. We recorded its visiting log from October 25th, 2006 to
February 6th, 2007 in the data format: time $t_i$/user ID number
$u_i$/music ID number $m_i$, i.e. a user $u_i$ listened to a song
$m_i$ at time $t_i$. Users were distinguished by their IP addresses.
The total number of log we obtained is 2,136,149, the number of
different music recorded is 98,747 (mostly popular songs), and the
number of users recorded is 8472.

$Wealink$ is a large social networking site in China whose users are
mostly professionals, typically businessmen and office clerks. Each
registered user has a profile, including his/her list of friends.
For privacy reasons, the data, logged from May 11th, 2005 (the
inception day for the Internet community) to August 22nd, 2007,
include only each user's ID and list of friends, and the time of
sending and accepting friendship invitations. The finial data format
is time $t_i$/user ID number $u_i$/user ID number $v_i$/flag $s_i$.
$s_i$ can take value of 0 or 1. $s_i=0$ indicates that at time $t_i$
a user $u_i$ invited another user $v_i$ to be his/her friend while
$s_i=1$ indicates that at time $t_i$ user $u_i$ accepted user
$v_i$'s invitation. During our data collection period, there are
273,395 sent invitations and more than 99.9\% have been accepted.
The total number of users recorded is $223,482$. Like most social
networking sites, in $Wealink$, only when the sent friendship
invitations are accepted, can the inviters and receivers become
online friends. We can measure users' activity and popularity in
terms of their numbers of sent and received invitations.

In $Comic$ individual activity and popularity can be well described
by stretched exponential distribution, which is shown in Fig. 3.
While in Fig. 4, the distribution of users' activity and popularity
in $Wealink$ has a power law tail.

\begin{figure}
  \centerline{\includegraphics[width=5.5in]{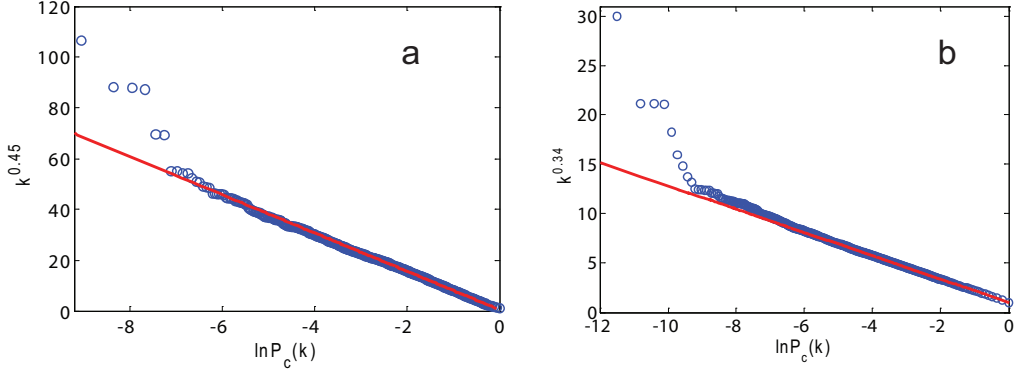}}
  \caption{The CCDFs for users' activity (a) and music's popularity (b)
in $Comic$. A stretched exponential distribution will show a
straight line if we use $\ln P_{\rm{c}} (k)$ as $x$-axis and $k^c$
as $y$-axis. The solid lines represent the fitted lines.}
\end{figure}

\begin{figure}
  \centerline{\includegraphics[width=4in]{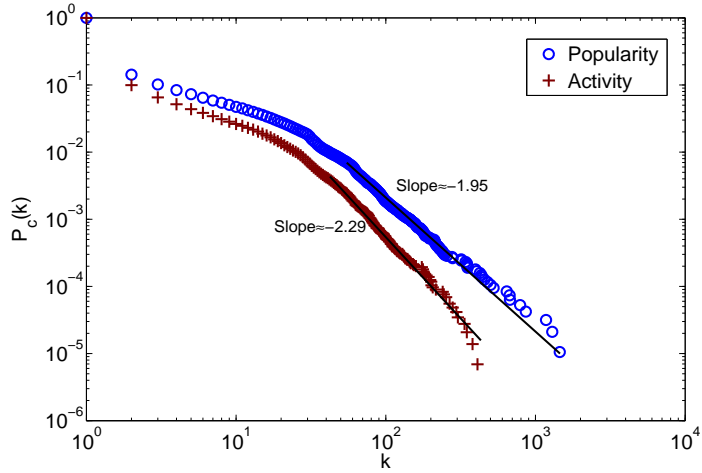}}
  \caption{The CCDFs of users' activity and popularity in $Wealink$.
Both distributions have a power law tail.}
\end{figure}

We can compare the distributions of individual activity and
popularity in real data with the predicted by the stochastic model.
Fig. 5 shows the $\kappa $ versus $k$ for music and users in
$Comic$. For these two cases, the sub-linear preferential selection
hypothesis can offer a good approximation. The values of $\beta$ for
users and music are approximately 0.61 and 0.79, respectively. For
the CCDF of users' activity, our model gives $c=1-\beta \approx
0.39$ and the empirical distribution in Fig. 3 gives $c \approx
0.45$, while for music's popularity, our model gives $c \approx
0.21$ and the empirical distribution gives $c \approx 0.34$. Fig. 6
shows the $\kappa $ versus $k$ for users in $Wealink$. Approximately
$\beta
 \approx 1$ for users' activity and popularity, indicating
linear preference. The appearance probabilities $\alpha$ of new
users in the time-ordered lists of users of sending and receiving
invitations are 0.53 and 0.35, respectively. For the CCDF of users'
activity, the model gives $\nu=1/(1-\alpha)=2.13$, while for users'
popularity the model gives $\nu=1.54$. The power law exponents
achieve proper agreement with the empirical results in Fig. 4.

\begin{figure}
  \centerline{\includegraphics[width=5.5in]{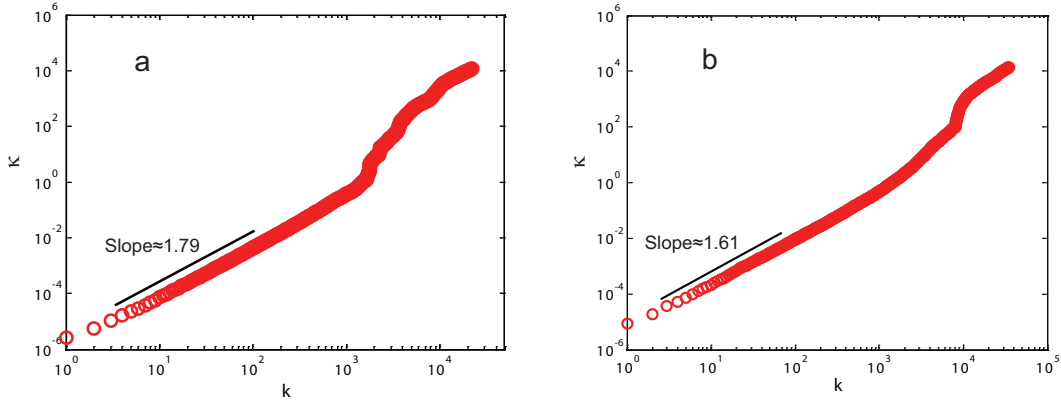}}
  \caption{$\kappa $ versus $k$ for music (a) and users (b) in $Comic$.}
\end{figure}

\begin{figure}
  \centerline{\includegraphics[width=4in]{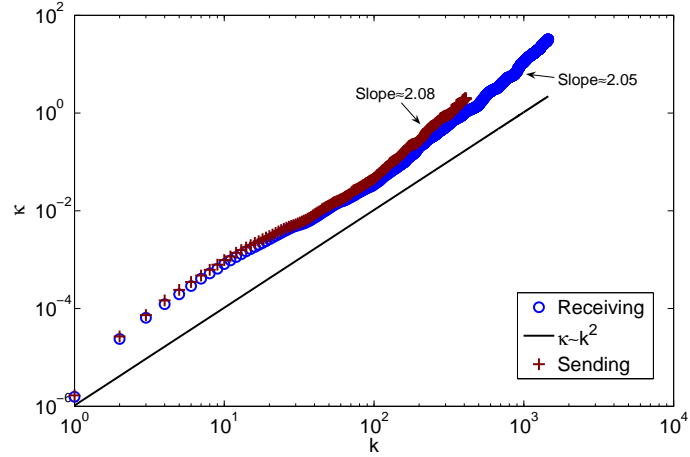}}
  \caption{Testing preferential selection for the users of sending
invitations and receiving invitations in \emph{Wealink}.}
\end{figure}

%\section{Discussion}
Fig. 7 shows the growth of the numbers of different users/music $N$
in $Comic$ and users $N$ in $Wealink$ with $T$. The traditional
assumption, as applied in the previous deduction, is that the
generation probability of new individuals is a constant value, i.e.
$N(T)\propto T$. However as shown in Fig. 7, the hypothesis is
unrealistic to some extent. For the $Wealink$ users, approximatively
the slopes for senders and receivers are 1.09 and 0.97,
respectively, however for the users/music in $Comic$, the growth
lines show several segments with different slopes. In some cases the
number of distinct items $N$ introduced by users after $T$
assignments can grow approximately as $N(T) \propto T^\gamma  $ with
$\gamma <1$. When dealing with the evolution of the number of
attributes pertaining to some collection of objects, this sub-linear
growth is generally referred to as Heaps' law [16]. As an example,
sub-linear behavior has been observed in the growth of vocabulary
size in texts, i.e. in the number of different words in a text as a
function of the total number of words observed while scanning
through it. For the case of English corpora, vocabulary growth
exponents in the range ${\rm{0}}{\rm{.4  < }}\gamma {\rm{ <
0}}{\rm{.6}}$ have been reported [17].

\begin{figure}
  \centerline{\includegraphics[width=4in]{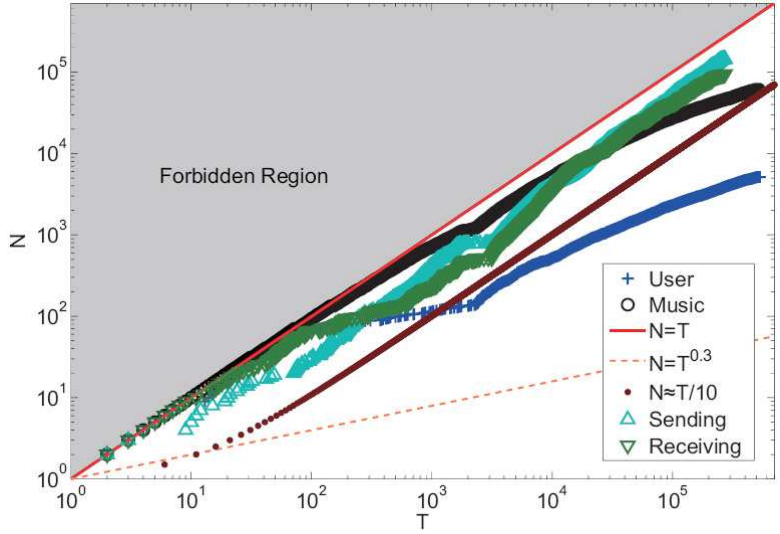}}
  \caption{Growth patterns of the numbers of different users/music $N$
in $Comic$ and users $N$ in $Wealink$ with $T$. For comparison the
approximately linear growth $N=(T+9)/10$($N \approx T/10$ for large
$T$) and sub-linear growth $N = T^{0.3} $ are also shown. Note that
growth curves are only feasible under $N=T$.}
\end{figure}

The rate at which new items appear at time $T$ scales as
${\rm{d}}N(T)/{\rm{d}}T \sim T^{\gamma  - 1} $. That is, new items
appear less and less frequently, with the invention rate of new
items monotonically decreasing towards zero. The approach to zero is
however so slow that the cumulated number of items, asymptotically,
does not converge to a constant value but is unbounded - assuming
the observed trend stays valid.

Different users or items with distinct activity or popularity may
have quite different $\gamma $. Recent research on the collaborative
tagging system $del.icio.us$ reveals that for less and less popular
resources being bookmarked, the distribution of growth exponent
$P(\gamma )$ of distinct tags gets broader and its peak shifts
towards higher values of $\gamma $, indicating that the growth
behavior is becoming more and more linear [3].

Table 1 summarizes the probability distributions of popularity and
activity for different patterns of growth and preferential selection
which can appear in real life. For sub-linear growth and linear
preference, the recent research shows that when the rate at which
new items appear $\dot N(T) \ll 1$, the distribution can be
approximately viewed as a power law $P(k)  \sim k^{ - 1 - \gamma } $
[18]. For sub-linear growth and sub-linear preference, unfortunately
the analysis for probability distribution can lead to a rather
intractable relation whose analytical solution is hard to obtain.
Qualitatively in this case the distribution is still a fat-tailed
one intermediate between exponential and power law. For some
sub-linear growth exponent, the distribution resulted from
sub-linear preference will be more homogeneous than that (power law)
resulted from linear preference; while for some sub-linear
preference exponent, the distribution resulted from sub-linear
growth will be more heterogeneous than that (stretched exponential
distribution) resulted from linear growth.

\begin{table}
\caption{Probability distributions of popularity and activity for
different patterns of growth and preference.}
\begin{center}
\begin{tabular}{lll}
\hline $ $  & Linear growth   &   Sub-linear growth\\  \hline
Linear preference   & $P(k)  \sim k^{ - \left( {1 + \frac{1}{{1 - \alpha }}} \right)} $ & $P(k)  \sim k^{ - 1 - \gamma } $\\
Sub-linear preference   & $P(k)  \sim k^{ - \beta }  \cdot \exp \left( { - \frac{{k^{1 - \beta }  - 1}}{{1 - \beta }} \cdot \frac{\mu }{{1 - \alpha }}} \right)$ & Fat tail\\
\hline
\end{tabular}
\end{center}
\end{table}

The distributions of individual popularity and activity in many
online social systems can follow generic heavy-tailed ones,
unnecessarily power law [19-24]. Several aspects of the underlying
intricate dynamics may be responsible for the feature. Except
sub-linear preference discussed above, another possible origin is
the memory effect, that is, newly appeared individuals will appear
more frequently than old ones. For example web users tend to listen
to recently added music or apply recently added tags more frequently
than old ones, which may be equivalent to the ageing effect of
individuals. The popularity or activity of an entity will inevitably
undergo a decaying process. Users become less active and items
become less attractive over the time [25-29].

According to growth and preference characteristic, it is possible to
predict the amount that would be devoted over time to given ones by
measuring the data at an early time. However the method does not
consider the semantics of popularity and why some items become more
popular than others [30]. That is, popularity prediction in the
presence of a large table of item assignments can essentially be
made based on the observed early time series, while semantic
analysis of content may be more useful when no early click-through
information is known. Semantic attraction can lead to the initial
prevalence of items and subsequent preferential selection
strengthens the popularity.

\section*{Acknowledgments}
We thank the anonymous reviewers for their constructive remarks and
suggestions which helped us to improve the quality of the manuscript
to a great extent. This work was partly supported by the NSF of PRC
under Grant No. 60674045.


\begin{thebibliography}{10}

\bibitem{1.} R. Lambiotte, M. Ausloos, Phys. Rev. E 72 (2005) 066107.

\bibitem{2.} S. Sinha, R. K. Pan, How a `hit' is born: The emergence of popularity from the dynamics of collective
choice, in: B. K. Chakrabarti, A. Chakraborti, A. Chatterjee (Eds.),
Econophysics and Sociophysics: Trends and Perspectives, Wiley-VCH,
Berlin, 2006, pp. 417-447.

\bibitem{3.} C. Cattuto, A. Baldassarri, V. D. P. Servedio, V.
Loreto, arXiv:0704.3316.

\bibitem{4.} C. Cattuto, V. Loreto, L. Pietronero, Proc. Natl. Acad. Sci. U. S. A. 104
(2007) 1461-1464.

\bibitem{5.} H. B. Hu, D. Y. Han, Physica A 387 (2008) 5916-5921.

\bibitem{6.} H. Hu, X. Wang, Physics Letters A 373 (2009) 1105-1110.

\bibitem{7.} A. Capocci, G. Caldarelli, J. Phys. A: Math. Gen. 41
(2008) 224016.

\bibitem{8.} F. Benevenuto, F. Duarte, T. Rodrigues, V. Almeida, J. Almeida, K.
Ross, arXiv:0804.4865.

\bibitem{9.} T. Zhou, H. A. T. Kiet, B. J. Kim, B. H. Wang, P. Holme, Europhys. Lett. 82
(2008) 28002.

\bibitem{10.} G. U. Yule, Phil. Trans. R. Soc. Lond. B 213 (1925)
21-87.

\bibitem{11.} H. A. Simon, Biometrika 42 (1955) 425-440.

\bibitem{12.} A. L. Barab\'{a}si, H. Jeong, Z. N\'{e}da, E. Ravasz, A.
Schubert, T. Vicsek, Physica A 311 (2002) 590-614.

\bibitem{13.} H. Jeong, Z. N\'{e}da, A. L. Barab\'{a}si, Europhys. Lett. 61 (2003) 567-572.

\bibitem{14.} P. L. Krapivsky, S. Redner, F. Leyvraz, Phys. Rev. Lett. 85 (2000) 4629-4632.

\bibitem{15.} B. Freiesleben de Blasio, \AA. Svensson, F. Liljeros, Proc. Natl. Acad. Sci. U. S. A. 104
(2007) 10762-10767.

\bibitem{16.} H. S. Heaps, Information Retrieval: Computational and Theoretical Aspects, Academic
Press, New York, 1978.

\bibitem{17.} D. Harman, Overview of the third text retrieval
conference, in: D. K. Harman (Eds.), Proc. Third Text REtrieval
Conference (TREC-3), NIST Special Publication 500-226, 1995, pp.
1-20.

\bibitem{18.} D. H. Zanette, M. A. Montemurro, J. Quant. Linguist. 12
(2005) 29-40.

\bibitem{19.} S. Whittaker, L. Terveen, W. Hill, L. Cherny, The dynamics of mass
interaction, in: Proc. ACM Conf. Computer-Supported Cooperative
Work, ACM Press, New York, 1998, pp. 257-264.

\bibitem{20.} P. Holme, C. R. Edling, F. Liljeros, Social Networks 26
(2004) 155-174.

\bibitem{21.} S. A. Golder, D. Wilkinson, B. A. Huberman, arXiv:cs/0611137.

\bibitem{22.} G. Ghoshal, P. Holme, Physica A 364 (2006) 603-609.

\bibitem{23.} Y. Y. Ahn, S. Han, H. Kwak, S. Moon, H. Jeong, Analysis of topological characteristics of huge online social networking
services, in: Proc. 16th Int. World Wide Web Conf., ACM Press, New
York, 2007, pp. 835-844.

\bibitem{24.} J. Leskovec, E. Horvitz, Planetary-scale views on a large Instant-Messaging
network, in: Proc. 17th Int. World Wide Web Conf., ACM Press, New
York, 2008, pp. 915-924.

\bibitem{25.} S. N. Dorogovtsev, J. F. F. Mendes, Phys. Rev. E 62
(2000) 1842.

\bibitem{26.} F. Wu, B. A. Huberman, Proc. Natl. Acad. Sci. U. S. A.
104 (2007) 17599-17601.

\bibitem{27.} A. Grabowski, N. Kruszewska, R. A. Kosi\'{n}ski, Eur. Phys. J. B 66
(2008) 107-113.

\bibitem{28.} Z. K. Zhang, L. L\"{u}, J. G. Liu, T. Zhou, Eur. Phys. J. B 66 (2008) 557-561.

\bibitem{29.} R. Crane, D. Sornette, Proc. Natl. Acad. Sci. U. S. A. 105
(2008) 15649-15653.

\bibitem{30.} G. Szabo, B. A. Huberman, arXiv:0811.0405.

\end{thebibliography}
\end{document}